\documentstyle [prl,aps,multicol,epsf]{revtex}
\begin{document}
\draft
\title{Polarized Brillouin Scattering in Salol: 
 Effects of Rotation-Translation Coupling    }
\author{  H. P. Zhang, A Brodin$^{*}$, H.C. Barshilia$^{\dag}$, G.Q. 
Shen$^{\ddag}$,  and H.Z. Cummins} 
\address {Physics Department,
City College of the City University of New York, New York, NY 10031}
\author{R.M. Pick}
\address {UFR 925, Universite P. \& M. Curie, Paris, France}
\date{\today}
\maketitle
\begin{abstract}

We have studied the $90^{\circ}$ VV-polarized Brillouin scattering  spectrum of the molecular
glassformer salol  and observed a previously unnoticed VV-dip feature at low frequencies
for temperatures above 300K. This new feature is a consequence of rotation-translation
coupling, as recently predicted by Pick, Franosch {\it et al} [Eur. Phys. J. B {\bf 31},  217, 229
(2003)], who showed its relationship to the Rytov dip that occurs in the corresponding VH
spectrum. The analysis of the spectra shows good agreement with the theoretical
predictions.

\end{abstract}

\begin{multicols}{2}
Thirty five years ago, the broad quasielastic rotational line in the depolarized
$I_{VH}^{90}(\omega)$ light-scattering  spectra of liquids composed of strongly anisotropic
molecules was found to have a  previously unknown feature \cite{Starunov67}, \cite{Stegeman68}.
With high resolution, at temperatures above the melting point, a narrow central dip  usually
designated as the "Rytov dip" \cite{Rytov58} is observed. For liquids that can
be supercooled, with decreasing temperature the Rytov dip disappears, the rotational line
narrows, and new spectral features develop symmetrically on the wings and, as
the glass transition temperature is approached, sharpen to become the TA
Brillouin components.

The origin of  the Rytov dip was shown by Andersen and Pecora \cite{Andersen71}, \cite{Berne76}
and  others \cite{Others71} to  be rotation-translation [RT] coupling, the same phenomenon
responsible for  flow birefringence. While the rotational spectral line is generated by
orientational  fluctuations, the dynamics of these fluctuations are modified by coupling of
orientation to  the transverse velocity. 

To explain the evolution of the VH spectrum with temperature, the equations of coupled
translation-plus-rotation dynamics must be modified to include viscoelasticity, i.e.
memory functions must be introduced for the relevant transport properties. This synthesis was
first described in a series of papers by C.H. Wang and his coworkers who carried out 
light-scattering studies of a series of molecular glassforming liquids \cite{Wang80}. If memory
effects are ignored, Wang's phenomenological theory reduces to the 
Andersen-Pecora results which were obtained from a two-variable Zwanzig-Mori formalism with the
Markov approximation.

Dreyfus, Pick and their coworkers have carried out a new phenomenological analysis of
the VH spectrum which is more complete than previous theories \cite{Dreyfus98}. In their
most recent publication, this analysis was extended to the polarized $I_{VV}^{90}(\omega)$
spectrum, and it was found that this spectrum should also exhibit a  characteristic
signature of RT coupling \cite{Pick2003}. Franosch {\it et al} \cite{Franosch2003}
implemented a parallel Zwanzig-Mori analysis of both the VH and VV spectra with results that
agree with the phenomenological results of Dreyfus {\it et al} and, if memory effects are
ignored, again recover the VH results of Andersen and Pecora.

In order to test these new theoretical results and, in particular, to search for the 
predicted new effect of RT coupling in the VV spectrum, we have carried out an 
extensive light-scattering study of the much-studied molecular glassforming liquid 
salol ($T_{M}$=316K, $T_{G}$=218K). We collected three complete sets of
spectra,
$90^{o}$VV, $90^{o}$VH, and VH backscattering, at 23 temperatures between 380K and 210K.
A  detailed analysis of all the spectra will be reported in a forthcoming publication
\cite{Zhang2003}. In this communication we discuss the $90^{o}$ VV spectra and describe the new
feature we have observed that confirms the predictions of Pick, Franosch {\it et al}
\cite{Pick2003}, \cite{Franosch2003}. In contrast to previous analyses of VV spectra which
assumed simple addition of the rotational and density-fluctuation spectra, RT
coupling must be included in the analysis.

In the phenomenological theory, the conventional equations of
hydrodynamics are augmented with an additional non-hydrodynamic
variable  to describe the departure from isotropy of the average molecular
orientation  within a small volume element.  For linear
(or axially symmetric) molecules, if $P(\theta, \phi, \vec{r}, t)$ is the
probability of finding a molecule at $\vec{r}$ with its axis $\hat{u}$
pointing in the direction $(\theta, \phi)$, an appropriate variable is
the orientational density which forms a symmetric traceless second-rank tensor
$\bar{\bar{Q}}$:
\begin{equation}
Q_{ij}(\vec{r},t) = \int d \theta d \phi P 
(\theta, \phi, \vec{r}, t) \: [ \hat{u}_{i}(\theta, \phi)
 \hat{u}_{j} (\theta, \phi) -\frac{1}{3}\delta_{ij} ].
\end{equation}

For the equation of motion of $\bar{\bar{Q}}$, Dreyfus and Pick  used a 
damped harmonic oscillator equation

\begin{equation}
\frac{\partial^{2}}{\partial t^{2}} Q_{ij} = -\omega^{2}_{R} Q_{ij}
- \Gamma\otimes \frac{\partial}{\partial t} Q_{ij} +\Lambda'\mu\otimes\tau_{ij}  
\end{equation}
where $\otimes$ denotes convolution, $\omega_{R}$ is a librational frequency, $\Gamma(t)$ is
the orientational  friction function, and the last term incorporates the
 coupling of orientation to the shear rate $\tau_{ij}$. 

In the Zwanzig-Mori approach, the damped oscillator equation for Q is obtained only if both Q
and dQ/dt are included in the set of selected variables. (With Q alone, the equation of motion
for Q is  purely relaxational.) This result parallels the fact that both the density
fluctuation $\rho_{q}$ and its time derivative (or current) must be included  to obtain the
usual damped oscillator equation for the density-fluctuation (LA) modes.

The fluctuations of the dielectric tensor include two terms, one due to
density fluctuations and the other to orientational fluctuations:
\begin{equation}
\delta {\bar{\bar \epsilon}} (\vec{r},t) = a \delta \rho_{m} (r, t)
 \: {\bar{\bar I}} + b \, {\bar{\bar Q}} \, (r,t)   
\end{equation}
Both terms in Eq. (3) as well as their cross products must be evaluated to find the complete
spectrum.  The resulting equation for the VV-polarized spectrum $I_{VV}^{90}(\omega)$ obtained
by Pick {\it et al} \cite{Pick2003}  (with some simplifying  approximations \cite{Zhang2003}) is
given by
\begin{equation}
I_{VV}(\omega)=\frac{I_{0}}{\omega}{\cal I}m
\left\{ \frac{4}{3}R(\omega)
 +\frac{q^{2}}{\rho_{m}}\Lambda~ [S+ \frac{2}{3}R(\omega)]^{2} 
P_{L}(\omega)\right\} \ \
\end{equation}
where the initial $\frac{4}{3}R(\omega)$ term is the pure orientational part of the spectrum
(which can be determined by  analysing  the VH backscattering spectra),  $\Lambda$ is the RT
coupling coefficient, and $P_{L}(\omega)$ is the longitudinal propagator:
\end{multicols}
\begin{equation}
P_{L}(\omega) =  \left\{ \omega^{2} - \omega_{0}^{2} -  i\omega\Gamma_{0} 
-\frac{q^{2}}{\rho_{m}} \left[ V \omega
\eta_{s}(\omega) - \frac{4}{3}\Lambda \frac{R^{2} (\omega)}{1 - R (\omega)}
\right]  \right\} ^{-1} 
\end{equation}

\begin{multicols}{2}
In Eq. (5), the longitudinal viscosity $\eta_{L}(\omega)$ does not appear explicitly because
we assume that it is proportional to the shear viscosity $\eta_{S}(\omega)$: 
$\eta_{L}(\omega) = V\eta_{S}(\omega)$, $ \omega_{0}$ is the bare LA frequency, and
$\Gamma_{0}$  is a constant damping term due to other processes.

The RT coupling mechanism enters twice. First in Eq. (4) where the
first term in the square brackets (S) represents the conventional light-scattering 
channel for LA modes via coupling of the dielectric constant to density fluctuations (the
first term in Eq. (3)), while the second term represents an additional light scattering
channel provided by  coupling of the LA mode to molecular orientation. Second in  Eq. (5)
where the final term gives the reduction of the shear  viscosity  by a factor proportional to
$\Lambda$. For salol, this reduction is $\sim$ 10\% and has only a minor effect on
the shape of the spectrum.

{\bf Experiments}:
 Salol samples for our light-scattering experiments  were prepared by triple vacuum 
distillation of phenyl salicylate purchased from Sigma Chemical
Company. The final distillation was made into cylindrical glass sample
cells which were then flame sealed under vacuum. The Brillouin scattering experiments were
performed with incident 5145 \AA  ~monomode laser excitation of $\sim
250$~mW; the spectra were measured with a Sandercock
6-pass tandem Fabry-Perot interferometer.

 In Fig. 1, the upper panel shows the 350K VV spectrum
(points) and the VH  backscattering spectrum (solid line) scaled to overlap the VV spectrum at
high  frequencies (30-50GHz). Numerical subtraction of the two spectra yields the 350K 
difference spectrum $I_{dif}(\omega)$ shown by the points in the lower panel. For frequencies
above $\simeq$ 15GHz, $I_{dif}(\omega)$ vanishes as expected. However, for 

\vbox{
\vspace{0 in}
\hbox{
\hspace{0in}
\epsfxsize 3.1in \epsfbox{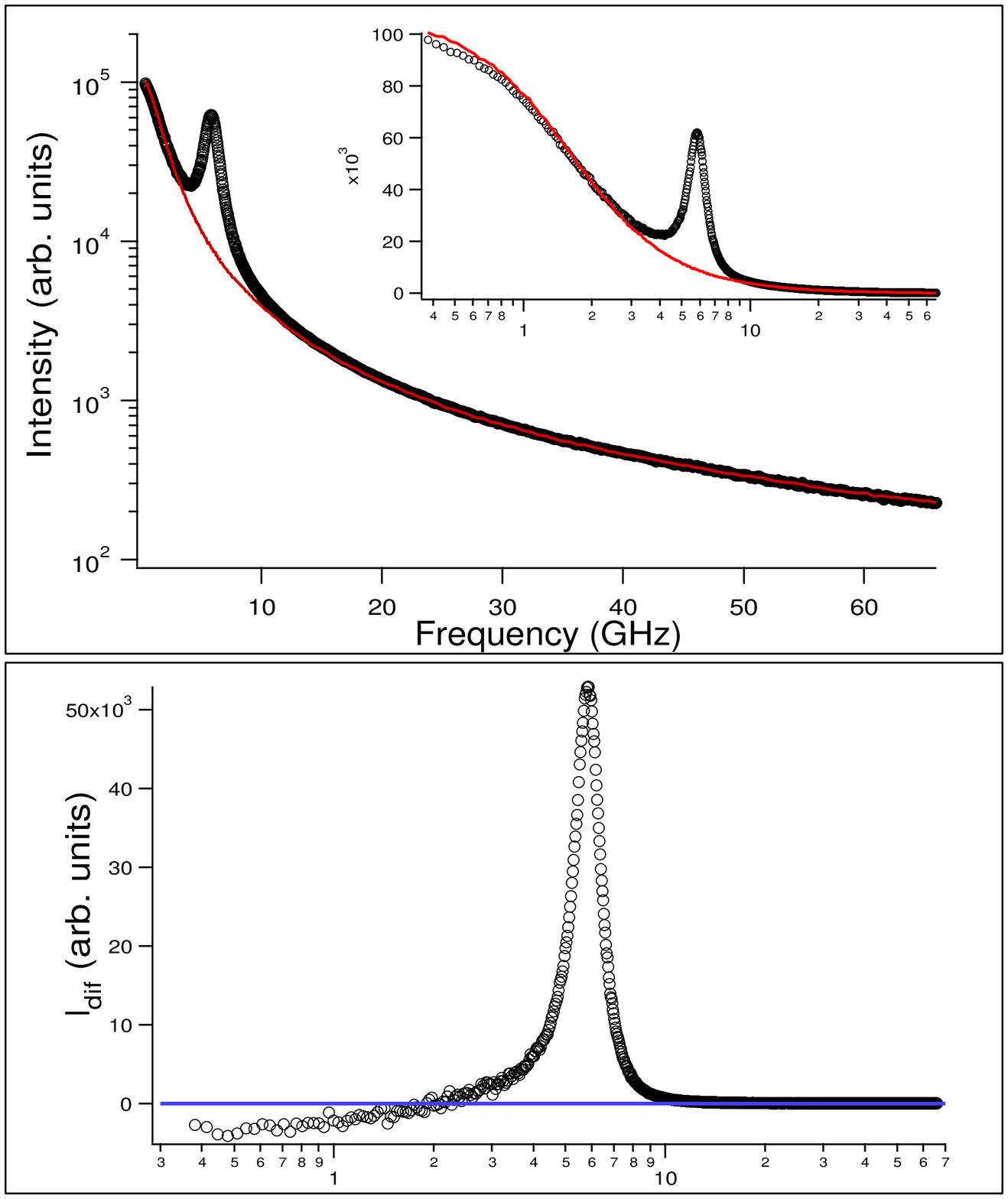}
}
\vspace{0.1in}
}
\refstepcounter{figure}
\parbox[b]{3.3in}{\baselineskip=12pt FIG.~\thefigure. Generating the $I_{ \rm dif}
(\omega)$ difference spectra.  Upper panel: $I_{VV}^{90}$ spectrum (points) and scaled
$I_{VH}$-back  spectrum (line) for $T=350$~K.  Inset: same, with linear $y$ and log $x$ scales. 
Lower panel: the difference spectrum obtained by subtracting the scaled
$I_{VH}$-back spectrum from the $I_{VV}^{90}$ spectrum. (The line is I = 0)
\vspace{0.10in}
}
\label{1}
\vspace{0.1in}
 
 frequencies
 $\leq 2GHz,  I_{dif}(\omega) < 0$. As we will see, it is this negative region which 
is the signature of the new RT coupling effect. $I_{dif}(\omega)$ is given by Eq. (4)
with the pure orientation term suppressed:
\begin{equation}
I_{dif}(\omega)=\frac{I_{0}}{\omega}{\cal I}m
\left\{ \frac{q^{2}}{\rho_{m}}\Lambda~ [S+ \frac{2}{3}R(\omega)]^{2}P_{L}(\omega)\right\} 
\end{equation}

In the fits, we modeled $R(\omega) = R^{0}H_{R}(\omega)$ and $\omega \eta_{S}(\omega) =
\eta_{S}^{0}H_{S}(\omega)$ with a "hybrid function"

\begin {equation}
H_{x}(\omega) = [1-(1+i\omega\tau_{x})^{-\beta_{x}} + i\omega p_{x}(\tau_{x}^{-1}
+i\omega)^{a-1}]  
\end {equation}
(where x = R or S) which combines the Cole-Davidson function at low frequencies with a
$\omega^{a}$ power-law at
 high frequencies to approximate the mode-coupling form in the frequency region of
the experiments \cite{Gotze92}. The parameters in 
$H_{R}(\omega)$ were determined from the VH backscattering spectra, and those in 
$H_{S}(\omega)$ and $\Lambda$ from the VH-$90^{o}$ spectra \cite{Zhang2003}. Both $\tau_{R}$
and $\tau_{S}$ increase rapidly with decreasing temperature with a nearly constant ratio.

{\bf Fits to the density-fluctuation-only model}: In analyzing the spectra, we initially
ignored  RT coupling and fit the difference spectra  with a  conventional
density-fluctuation-only model. If RT coupling is ignored,  Eqs.~(5) and ~(6) reduce to

\begin{equation}
I_{dif} (\omega) = \frac{A}{\omega} {\cal I}m [ P_{L}
(\omega) ]  
\end{equation}
where $A = I_{0}(q^{2}/\rho_{m})\Lambda S^{2}$, and

\begin{equation}
P_{L}(\omega) = \left[ \omega^{2} - \omega_{0}^{2} -
i \omega \Gamma_{0} - \Delta^{2} H_{S} (\omega) \right]^{-1}
\end{equation}
where $H_{S}(\omega)$ is the hybrid function (Eq.~[7]) for the shear viscosity
$\eta_{S}(\omega) = \eta_{S}^{\circ} H_{S} (\omega) $   
and $\Delta ^{2} = (q^{2}/ \rho_{m}) V \eta_{S}^{\circ}$.

We  fit the difference spectra to Eqs.~(8) and (9), with
the parameters appearing in $H_{S}(\omega)$ and $\Lambda$ again fixed from the $I_{VH}^{90}$
fits.  This is a conventional analysis of
Brillouin scattering spectra similar to that carried out previously for 
many other materials (e.g. refs.~ \cite{Brodin2002}, \cite{Wiedersich2000}.)
In the fits, $A$, $\Delta^{2}$, and $\omega_{0}$ were the free parameters.  We 
found that good fits could be obtained with fixed
$\Gamma_{0}$ proportional to $T$, as was found for propylene carbonate \cite{Brodin2002}. 
The choice
$\Gamma_{0} = T/600$ was found to produce values of $\omega_{0}$
quite close to the values computed from ultrasonic data.  These "density-fluctuation-only" fits
are shown in Fig.~2 by broken lines.  While the fits
are generally excellent, for  temperatures above 280K the  low-frequency regions of the
fits become increasingly less satisfactory because the spectra become negative at low
frequencies while the theoretical fits are always positive. (Note that while the
difference spectra have low-frequency negative regions, {\newline}

\vbox{
\vspace{0.2 in}
\hbox{
\hspace{0in}
\epsfxsize 3.2in \epsfbox{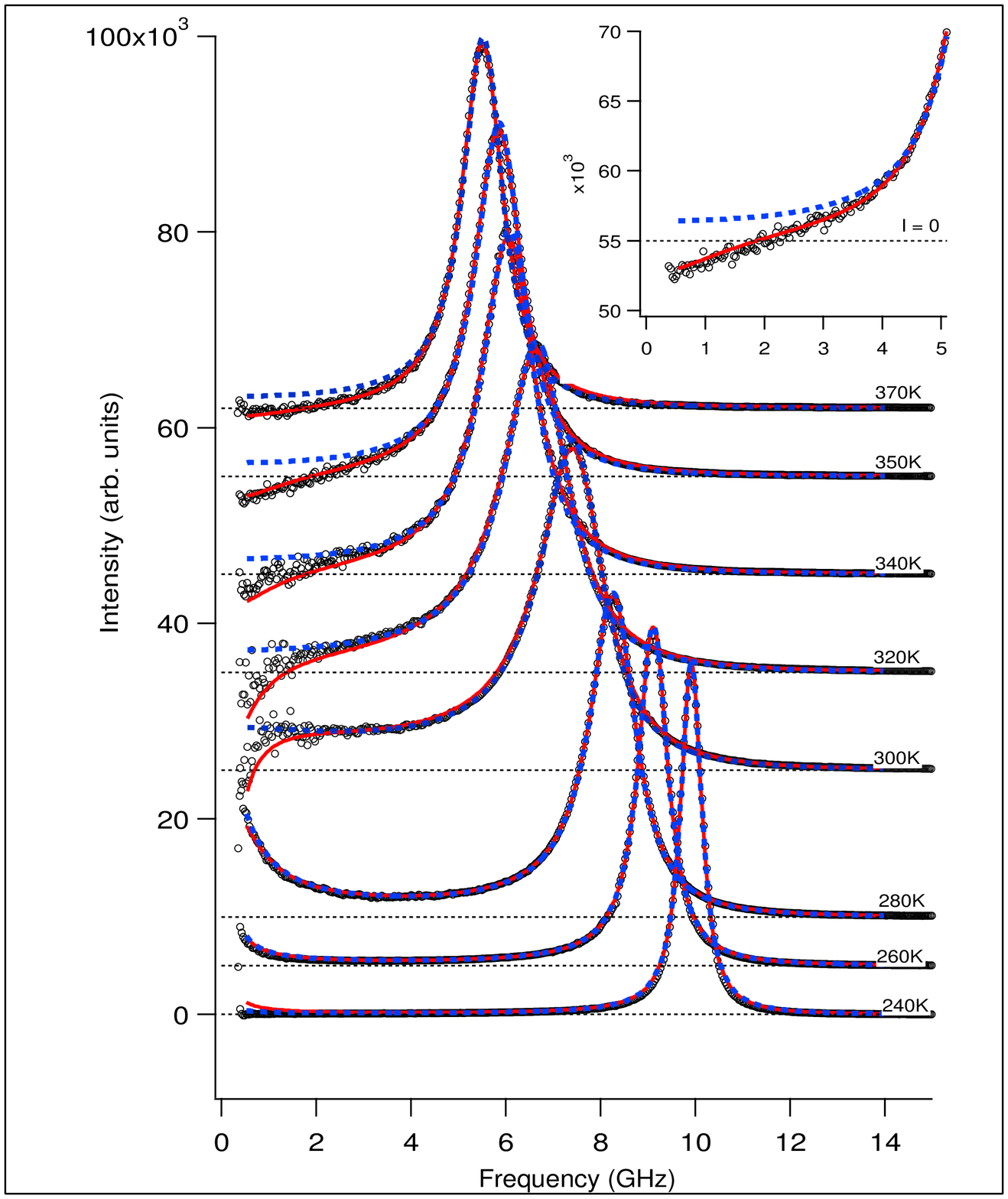}
}
\vspace{0.1in}
}
\refstepcounter{figure}
\parbox[b]{3.3in}{\baselineskip=12pt FIG.~\thefigure.
 Difference spectra for $T$ =240, 260, 280, 300, 320, 340, 350, and 370K. The
spectra have been scaled and shifted vertically for visibility. Solid lines: fits to the
full theory; broken lines: fits to the density-fluctuation-only model. The dotted lines
indicate the I=0 baseline for each spectrum. Inset: Low-frequency region for T = 350K.
\vspace{0.10in}
}
\label{1}
\vspace{0.1in}
the full $I_{VV}(\omega)$ spectra are everywhere positive).

 {\bf Fits to the full theory}:
Next, we fit the $I_{\rm dif}(\omega)$ spectra using the full theory of Eqs.~(5) and
(6) with $\omega_{0}$, V, and S as the free fitting
parameters. The fit results 
are shown in Fig. 2 where we have included both
the full fits (solid  lines) and the density-fluctuation-only fits (broken  lines).
At low temperatures the two fits are indistinguishable, showing that RT coupling is unimportant
in the frequency range of our spectra. At higher temperatures, the low-frequency region is fit
much better by the full theory than by the density-fluctuation-only theory,  showing that 
RT coupling, which is ignored in the conventional density-fluctuation-only analysis, is the
source of the negative region.

Referring to Fig. 4 in ref. \cite{Pick2003}, there is also a tendency for the full theory to
fall below the density-fluctuation-only fits to it at low frequencies when $\omega_{B}\tau \simeq
1$. There are differences between the lineshapes in that figure and ours, however, which we
attribute primarily to the use of a single relaxation time in their calculations, while for
salol,
 $\tau_{R} \sim 15\tau_{S}$. Because of this large ratio, there is a clear separation
between the Mountain mode visible as a positive feature at temperatures between 260 and
290K and the VV dip, visible for temperatures above 300K. The Mountain mode results from
coupling of strain to structural relaxation and, at low temperatures, is visible at
frequencies below the Brillouin peak. For $T\geq$ 300K, it is cut off at low frequencies
by the VV dip.

Physically, RT coupling enters because the uniaxial strain that characterizes
a longitudinal acoustic mode is a superposition of simple compression and shear strain.
In liquids composed of anisotropic molecules, the fluctuating longitudinal strain 
induces  a preferential
orientation of the molecules in the plane perpendicular to $\vec{q} = q\hat{z}$, adding to 
$\delta\epsilon_{yy}$ which determines the VV scattered intensity. Referring to the
equation of motion for Q (Eq. [2]), the steady-state solution gives $Q_{ij}$ proportional to
the strain rate $\tau_{ij}$. Therefore, for oscillatory LA waves  in the low-frequency limit
(where
$\omega\tau << 1$), if $\delta\rho_{ij}$ is real, then $\delta Q_{ij}$ is imaginary. In this limit,
the optical coupling function $C(\omega) = [S  + (2/3)R(\omega)]^{2}$ in Eq. [6] becomes

\begin{equation}
[S  + \frac{2}{3}R(\omega)]^{2}\rightarrow [S+i\omega\frac{2}{3}R_{0}\beta_{R}\tau_{R}]^{2}
\approx [S^{2}+i\omega\frac{4}{3} R_{0}\beta_{R}\tau_{R}] 
\end{equation}
and, from Eqs. [5] and [6] in this low-frequency limit,

\begin{equation}
I_{dif}(\omega) \rightarrow \frac{I_{0}q^{2}\Lambda}
{\rho_{m}\omega_{0}^{4}}[S^{2}\eta_{L}\frac{q^{2}}{\rho_{m}}- S\omega_{0}^{2} \frac{4}{3}
R_{0}\beta_{R}\tau_{R}] 
\end{equation}
The second term, being negative, reduces the intensity at low frequencies, but it will disappear
at frequencies $\omega > \tau_{R}^{-1}$ where Q can no longer follow the oscillating strain.
This reduction effect is similar to the Rytov dip where, for $\omega\tau_{R}<<1, R^{2}(\omega)$
is negative which adds a narrow negative component to the center of the broad rotational line.

We therefore conclude that the VV-dip phenomenon observed in this experiment is a
consequence of RT coupling as described in the analysis of references \cite{Pick2003} and
\cite{Franosch2003}. Finally, we note that there have been many previous Brillouin-scattering
studies of liquids of anisotropic molecules (including salol), where the VV-dip could have
been observed. The reason that it has not been reported previously is suggested by the
upper panel of Fig. 1 which shows that on a log plot, the low-frequency
$I_{VV}^{90}(\omega)$ spectrum is very close to the rotational
spectrum. It is only when the low-frequency region of the
{\em difference spectrum} is examined that the small low-frequency VV dip becomes
apparent. An extensive discussion of this conclusion together with the full underlying data
analysis can be found in Ref \cite{Zhang2003}.

We thank W. G\"{o}tze, T. Franosch, C. Dreyfus, M. Fuchs, and
C.H. Wang for helpful discussions.  This material is based upon work supported by
the National Science Foundation under grants DMR-9980370 and
DMR-0243471.

\end{multicols}
\end{document}